\documentstyle[amstex,12pt,epsfig]{article}

\newlength{\dinwidth}
\newlength{\dinmargin}
\begin{document}



\thispagestyle{empty} \vspace*{1cm} \rightline{Napoli DSF-T-25/2002} %
\rightline{INFN-NA-25/2002} \vspace*{2cm}

\begin{center}
{\LARGE Twisted CFT and bilayer Quantum Hall systems }

{\LARGE in the presence of an impurity}

{\LARGE \ }

\vspace{8mm}

{\large Gerardo Cristofano\footnote{{\large {\footnotesize Dipartimento di
Scienze Fisiche,}{\it \ {\footnotesize Universit\'{a} di Napoli ``Federico II''
\newline and INFN, Sezione di Napoli}-}{\small Via Cintia - Compl.
universitario M. Sant'Angelo - 80126 Napoli, Italy}}}, Vincenzo
Marotta\footnotemark[1], }

{\large Adele Naddeo\footnote{{\large {\footnotesize Dipartimento di Scienze
Fisiche,}{\it \ {\footnotesize Universit\'{a} di Napoli ``Federico II''
\newline
and INFM, Unit\`{a} di Napoli}-}{\small Via Cintia - Compl. universitario M.
Sant'Angelo - 80126 Napoli, Italy}}} }

{\small \ }

{\bf Abstract\\[0pt]
}
\end{center}

\begin{quotation}
We identify the impurity interactions of the recently proposed CFT
description of a bilayer Quantum Hall system at filling $\nu =\frac{m}{pm+2}$
\cite{cgm2}. Such a CFT is obtained by $m$-reduction on the one layer
system, with a resulting pairing symmetry and presence of quasi-holes. For
the $m=2$ case boundary terms are shown to describe an impurity interaction
which allows for a localized tunnel of the Kondo problem type.The presence
of an anomalous fixed point is evidenced at finite coupling which is
unstable with respect to unbalance and flows to a vacuum state with no
quasi-holes.

\vspace*{0.5cm}



{\footnotesize Work supported in part by the European Communities Human
Potential}

{\footnotesize Program under contract HPRN-CT-2000-00131 Quantum
Spacetime\newpage }\baselineskip=18pt \setcounter{page}{2}
\end{quotation}

\section{Introduction}

Recently a powerful technique, the $m$-reduction procedure \cite{cgm1}, was
successfully employed in \cite{cgm2} in order to construct an effective
Conformal Field Theory (CFT)\ for a Quantum Hall system of two interacting
layers at total filling $\nu =\frac{1}{p+1}$. In such a description, the
Twisted Model (TM), the neutral current originates from a point-like
interaction between the two layers, which can be attributed to the presence
of a localized impurity (twist) on the edge. Localized (static) impurities,
even though do not change the central charge of the CFT, strongly modify the
properties of the ground state and of the spectrum. In fact for any
impurities class the Hamiltonian contains a boundary interaction term, which
not only determines the ground state energy shift but also gives information
on its stability ($g$-theorem) so that different universality classes exist
for a given filling. For a bilayer system, which exhibits interesting
phenomena such as interlayer phase coherence \cite{Sp-}\cite{Girvin}, a few
universality classes are known, corresponding to different values of the
relevant parameters as the distance between the layers and the symmetric
tunneling strength. In a CFT approach they differ in the description of the
neutral modes which can be given in terms of symplectic (Haldane-Rezayi
model \cite{Haldan-R}), Dirac (Halperin model \cite{Halperin} (H)) or Ising$%
^{2}$ fermions (TM). While the neutral degrees of freedom contribution to
the central charge is $c=-2$ for the first model, for the other two ones its
contribution is $c=1$ and the only difference is in their symmetry. They
relate to the Moore-Read (MR) Pfaffian model in which the fundamental
particles appear as p-wave BCS like paired states. Pairing symmetry also
implies non-Abelian statistics as well as the presence of quasi-holes.

In the TM model the ``bulk'' degrees of freedom for the bilayer system are
described in terms of a free compactified boson $X$ with a $U(1)$ symmetry,
for the charged sector, and a $Z_{2}$-twisted $\phi $ field for the neutral
ones, at compactification radius $R^{2}=1$ (see \cite{cgm2} for details). On
the other hand boundary conformal field theory has been successfully applied
to solve quantum impurity problems. By folding a two dimensional system at
the defect line, associated with the impurity, the problem is mapped to a
system with a boundary. In particular, in \cite{Affleck}, starting from the $%
c=\frac{1}{2}$ Ising CFT, by folding, an (Ising)$^{2}$ $c=1$ CFT is
obtained, which can be also described as a $Z_{2}$ orbifold of a free boson
at compactification radius $R^{2}=1$. In such a boundary CFT, apart for a
continuous boundary there are eight mixed boundaries, which are explicitly
realized by the Ishibashi states and are fixed at the two ends of the
orbifold $S^{1}/Z_{2}$ line.

That appears then a natural framework for studying the stability of the
degrees of freedom content of our model (the TM) in terms of boundary
states. It is the aim of this letter to express the full set of boundary
operators corresponding to the primary fields found in ref. \cite{cgm4}.
Indeed we can introduce the chiral partition function $Z_{AB}=<A|e^{-LH}|B>$%
, where $A$, $B$ are boundary states (BS), and express it as a superposition
of characters of the free bulk CFT. It turns out that the TM characters are
in a simple correspondence with the nine possible boundaries of the CFT
given in \cite{Affleck}. Furthermore, to convince the reader of the
relevance of the TM for the description of a system of highly correlated
electrons in the presence of magnetic impurities, we also trace an explicit
correspondence between the anisotropic two channel one impurity Kondo
problem and the TM (for $p=0$).

The paper is organized as follows:

In sec. 2 we shortly review the construction of the effective CFT of the TM
and its degrees of freedom content in terms of the allowed characters.

In sec. 3 a brief review of the boundary CFT is presented, the Ishibashi
states and Cardy consistency conditions recalled and generalized to the
orbifold case. The TM characters are then expressed in terms of boundary
states.

In sec. 4 the $g$-stability of the different vacua is analyzed in the
framework of the Kondo problem and the identification of the non trivial
fixed point (in the Toulouse limit) with the twisted ground state shown.

In sec. 5 a quick summary with comments is presented.

\section{The twisted model}

In this section we briefly review the $m$-reduction procedure for the
special $m=2$ case (see ref.\cite{cgm2} for the general case), since we are
interested in a system consisting of two parallel layers of 2D electrons gas
in a strong perpendicular magnetic field. The filling factor $\nu ^{(a)}=%
\frac{1}{2p+2}$ is the same for the two $a=1$, $2$ layers while the total
filling is $\nu =\nu ^{(1)}+\nu ^{(2)}=\frac{1}{p+1}$. For $p=0$ ($p=1$) it
describes the $220$ ($331$) Halperin state \cite{Halperin}. The CFT
description for such a system can be given in terms of two compactified
chiral bosons $Q^{(a)}$ with central charge $c=2$.

In order to construct the field $Q^{(a)}$ for the TM, let us start from the
bosonic ``Laughlin'' filling $\nu =1/2(p+1)$, described by a CFT with $c=1$
in terms of a scalar chiral field $Q$ compactified on a circle with radius $%
R^{2}=1/\nu =2(p+1)$ (or its dual $R^{2}=2/(p+1)$). It is explicitly given
by:
\begin{equation}
Q(z)=q-i\,p\,lnz+\sum_{n\neq 0}\frac{a_{n}}{n}z^{-n}  \label{modes}
\end{equation}
with $a_{n}$, $q$ and $p$ satisfying the commutation relations $\left[
a_{n},a_{n^{\prime }}\right] =n\delta _{n,n^{\prime }}$ and $\left[ q,p%
\right] =i$.

The $U(1)$ current $J(z)$ is given by $J(z)=i\partial _{z}Q(z)$ and the
primary fields are expressed in terms of the vertex operators $U^{\alpha
}(z)=:e^{i\alpha Q(z)}:$ with $\alpha ^{2}=1,...,2(p+1)$ and conformal
dimension $h=\frac{\alpha ^{2}}{2}$.

From such a CFT (mother theory), using the $m$-reduction procedure, which
consists in considering the subalgebra generated only by the modes in eq.(%
\ref{modes}) which are a multiple of an integer $m$, we get a $c=m$ orbifold
CFT (daughter theory, i.e. the TM) which describes the LLL dynamics. Then
the fields in the mother CFT can be organized into components which have
well defined transformation properties under the discrete $Z_{m}$ (twist)
group, which is a symmetry of the TM. By using the mapping $z\rightarrow
z^{1/m}$ and by making the identifications $a_{nm+l}\longrightarrow \sqrt{m}%
a_{n+l/m}$, $q\longrightarrow \frac{1}{\sqrt{m}}q$ (see ref. \cite{VM}) the $%
c=m$ CFT (daughter theory) is obtained.

Its primary fields content, for the special $m=2$ case, can be expressed in
terms of:

1) the $Z_{2}$-invariant scalar field $X(z)$, given by
\begin{equation}
X(z)=\frac{1}{2}\left( Q^{(1)}(z)+Q^{(2)}(-z)\right)  \label{X}
\end{equation}
corresponding to a boson compactified on a circle with radius $R_{X}$ now
equal to $R_{X}^{2}=R^{2}/2=p+1$, which describes the electrically charged
sector of the new filling;

2) the twisted field
\begin{equation}
\phi (z)=\frac{1}{2}\left( Q^{(1)}(z)-Q^{(2)}(-z)\right)  \label{phi}
\end{equation}
which satisfies the twisted boundary conditions $\phi (e^{i\pi }z)=-\phi (z)$
and describes the neutral sector \cite{cgm2}. Notice that its
compactification radius is $R_{\phi }^{2}=1$ independent on the flux $p$.

The chiral fields $Q^{(a)}$, defined on a single layer $a=1$, $2$, due to
the boundary conditions imposed upon them by the orbifold construction, can
be thought as components of a unique ``boson'' defined on a double covering
of the disc (layer) ($z_{i}^{(1)}=-z_{i}^{(2)}=z_{i}$). As a consequence of
such a construction the two layers system becomes equivalent to one-layer
QHF (in contrast with the Halperin model in which they appear independent)
and the $X$ and $\phi $ fields defined in eqs. (\ref{X}) and (\ref{phi})
diagonalize the interlayer interaction. In particular, the $X$ field carries
the total charge with velocity v$_{X}$ while $\phi $ carries the charge
difference of the two edges with velocity $v_{\phi }$ i.e. no charge, being
the number of electrons the same for each layer (balanced system) \cite
{priadko}.

Correspondingly the Virasoro generator splits into the two terms \cite{VM}:
\begin{equation}
T_{X}(z)=-{\frac{1}{2}}\left( \partial _{z}X(z)\right) ^{2}  \label{STRESSX}
\end{equation}
and
\begin{equation}
T_{\phi }(z)=-\frac{1}{4}\left( \partial _{z}\phi (z)\right) ^{2}+\frac{1}{%
16z^{2}}  \label{STRESSFI}
\end{equation}
each contributing with $c=1$ to the central charge. The primary fields are
composite operators and, on the torus, they are described in terms of the
conformal blocks of the MR and the Ising model \cite{cgm4}. The MR
characters $\chi _{(\lambda ,s)}^{MR}$ with $\lambda =0,...2$ and $s=0,...,p$%
,\ are explicitly given by:
\begin{eqnarray}
\chi _{(0,s)}^{MR}(w|\tau ) &=&\chi _{0}(\tau )K_{2s}\left( w|\tau \right)
+\chi _{\frac{1}{2}}(\tau )K_{2(p+s)+2}\left( w|\tau \right) \\
\chi _{(1,s)}^{MR}(w|\tau ) &=&\chi _{\frac{1}{16}}(\tau )\left(
K_{2s+1}\left( w|\tau \right) +K_{2(p+s)+3}\left( w|\tau \right) \right) \\
\chi _{(2,s)}^{MR}(w|\tau ) &=&\chi _{\frac{1}{2}}(\tau )K_{2s}\left( w|\tau
\right) +\chi _{0}(\tau )K_{2(p+s)+2}\left( w|\tau \right).
\end{eqnarray}
They represent the field content of the $Z_{2}$ invariant $c=3/2$ \ CFT \cite
{M-R} with a charged component ($K_{\alpha }(w|\tau )=\frac{1}{\eta (\tau )}%
\Theta \left[
\begin{array}{c}
\frac{\alpha }{4\left( p+1\right) } \\
0
\end{array}
\right] \left( 2\left( p+1\right) w|4\left( p+1\right) \tau \right) $) and a
neutral component ($\chi _{\beta }$, the conformal blocks of the Ising
Model).

The characters of the twisted sector are given by:
\begin{eqnarray}
\chi _{(0,s)}^{+}(w|\tau ) &=&\bar{\chi}_{\frac{1}{16}}\left( \chi
_{(0,s)}^{MR}(w|\tau )+\chi _{(2,s)}^{MR}(w|\tau )\right) \\
\chi _{(1,s)}^{+}(w|\tau ) &=&\left( \bar{\chi}_{0}+\bar{\chi}_{\frac{1}{2}%
}\right) \chi _{(1,s)}^{MR}(w|\tau )
\end{eqnarray}
which do not depend on the parity of $p$;
\begin{eqnarray}
\chi _{(0,s)}^{-}(w|\tau ) &=&\bar{\chi}_{\frac{1}{16}}\left( \chi
_{(0,s)}^{MR}(w|\tau )-\chi _{(2,s)}^{MR}(w|\tau )\right) \\
\chi _{(1,s)}^{-}(w|\tau ) &=&\left( \bar{\chi}_{0}-\bar{\chi}_{\frac{1}{2}%
}\right) \chi _{(1,s)}^{MR}(w|\tau )
\end{eqnarray}
for $p$ even, and
\begin{eqnarray}
\chi _{(0,s)}^{-}(w|\tau ) &=&\bar{\chi}_{\frac{1}{16}}\left( \chi _{0}-\chi
_{\frac{1}{2}}\right) \left( K_{2s}\left( w|\tau \right) +K_{2(p+s)+2}\left(
w|\tau \right) \right) \\
\chi _{(1,s)}^{-}(w|\tau ) &=&\chi _{\frac{1}{16}}\left( \bar{\chi}_{0}-\bar{%
\chi}_{\frac{1}{2}}\right) \left( K_{2s+1}\left( w|\tau \right)
-K_{2(p+s)+3}\left( w|\tau \right) \right)
\end{eqnarray}
for $p$ odd.

Notice that the last two characters are not present in the TM partition
function and that only the symmetric combinations $\chi _{(i,s)}^{+}$ can be
factorized in terms of the $c=\frac{3}{2}$ \ and $c=\frac{1}{2}$ theory.
That is a consequence of the parity selection rule ($m$-ality), which gives
a gluing condition for the charged and neutral excitations.

Furthermore the characters of the untwisted sector are given by:
\begin{eqnarray}
\tilde{\chi}_{(0,s)}^{+}(w|\tau ) &=&\bar{\chi}_{0}\chi _{(0,s)}^{MR}(w|\tau
)+\bar{\chi}_{\frac{1}{2}}\chi _{(2,s)}^{MR}(w|\tau )  \label{vacuum1} \\
\tilde{\chi}_{(1,s)}^{+}(w|\tau ) &=&\bar{\chi}_{0}\chi _{(2,s)}^{MR}(w|\tau
)+\bar{\chi}_{\frac{1}{2}}\chi _{(0,s)}^{MR}(w|\tau ) \\
\tilde{\chi}_{(0,s)}^{-}(w|\tau ) &=&\bar{\chi}_{0}\chi _{(0,s)}^{MR}(w|\tau
)-\bar{\chi}_{\frac{1}{2}}\chi _{(2,s)}^{MR}(w|\tau )  \label{vacuum2} \\
\tilde{\chi}_{(1,s)}^{-}(w|\tau ) &=&\bar{\chi}_{0}\chi _{(2,s)}^{MR}(w|\tau
)-\bar{\chi}_{\frac{1}{2}}\chi _{(0,s)}^{MR}(w|\tau ) \\
\tilde{\chi}_{(s)}(w|\tau ) &=&\bar{\chi}_{\frac{1}{16}}\chi
_{(1,s)}^{MR}(w|\tau )
\end{eqnarray}
We point out that the periodic sector of the TM describes a model for the
bilayer which was introduced by Ho in \cite{Ho}. In \cite{cgm2} it was
observed that this sector has degeneracy $3(p+1)$ instead of the $4(p+1)$
one of the Halperin model. The difference between the Halperin and the Ho
description relies in the different strength of the Coulomb interaction
between the electrons in the same layer with respect to that between
electrons in different ones. From the CFT point of view these effects modify
the neutral sector from a Dirac to an Ising$^{2}$ fermionic theory and it is
a consequence of the Abelian orbifold considered there. Moreover the CFT for
the Ho model, which was introduced to describe a continuous transition to a
Pfaffian MR state due to the degeneracy matching, is inconsistent without
the twisted sector of the TM (see \cite{cgm2}). The reduction of the degrees
of freedom is basically due to the $Z_{2}$ symmetry of the TM and is closely
related to its non-Abelian statistics. That seems to be a peculiarity of the
TM with respect to the unorbifolded H model. Furthermore the H model has an $%
U(1)$ symmetry which is explicitly broken in the TM and an observable phase
can be induced by the continuous Dirichlet BS introduced in the next section
in eq.(\ref{continous}). As a consequence a phase coherence between the
particles on the two layers takes place and a Josephson like effect is
possible \cite{nayak-Wilczek}.

\section{Boundary CFT and the twisted model}

Homogenous systems are in many cases idealizations of real systems, which in
general present disomogeneities or defects. Boundary Conformal Field Theory
(BCFT) has been developed in the last few years to solve such problems. In
particular an interesting relationship has been traced between the BCFT and
the Kondo effect, where the interaction energy of the conduction electrons
with the quantum impurity internal degrees of freedom results, at large
distances, into a boundary condition for the fields describing the electrons
in the bulk. Furthermore such a non trivial boundary condition in most cases
turns out to be conformally invariant. More recently BCFT has been applied
to the two dimensional Ising model with a defect line \cite{Affleck}. By
folding the system at the defect line, a mapping can be made to a boundary
problem and the BS can be identified in the $c=1$ $Z_{2}$ orbifold CFT of a
free boson. The construction of the BS goes basically through two
consistency requirements: the Ishibashi and the Cardy conditions \cite{Cardy}%
. In order to classify boundary conditions let us consider a cylinder with a
periodic direction of length $1/T$ and of length $L$ in the axis direction.
If we indicate with A, B the boundary conditions on the cylinder edges we
can evaluate the partition function $Z_{AB}$ for such a configuration in two
ways:

\begin{itemize}
\item  $Z_{AB}=tr\ e^{-\frac{H_{AB}}{T}}$- that is the cylinder surface may
be viewed as swept by an open string (with boundary conditions A, B at its
end points) which propagates in a closed loop of length $1/T$, $H_{AB}$
being its Hamiltonian depending on the boundary conditions A, B;

\item  $Z_{AB}=<A|e^{-LH}|B>$ - that is the closed string boundary state $|B>
$ propagates along L with a periodic Hamiltonian $H$ turning into a boundary
state $|A>$.
\end{itemize}

The corresponding expressions are given in terms of characters of the bulk
theory: $Z_{AB}=\sum_{ij}n_{AB}^{i}S_{i}^{j}\chi _{j}(q)$, where $S$ gives
the modular transformations of the characters $\chi $ and $n_{AB}^{i}$ is
the number of times the irreducible representation of highest weight $i$
appears in the spectrum of the crossed-channel Hamiltonian between the
boundary A and B.

The BS must satisfy the condition: $\left( L_{n}-\bar{L}_{-n}\right) |B>=0$,
required by conformal invariance or a stronger condition\ $\left( a_{n}\pm
\bar{a}_{-n}\right) |B>=0$ where $a_{n}$, $\bar{a}_{-n}$ are free
oscillators. The BS satisfying the above equation (Dirichlet for the $-$
sign, Neumann for the $+$ sign) are defined up to a normalization constant
which is fixed by the Cardy condition rising from equating $Z_{AB}$ in two
descriptions given before. The duality between the electric charge and the
magnetic flux also exchange Dirichlet and Neumann boundary conditions. The
folding procedure is used in the literature to map a problem with a defect
line (as a bulk property) into a boundary one, where the defect line appears
as a boundary state of a theory which is not anymore chiral and its fields
are defined in a reduced region which is one half of the original one. Our
approach in \cite{cgm2} is a chiral description of that, where the chiral $%
\phi $\ field defined in ($-L/2$, $L/2)$ describes both the left moving
component $\phi _{L}$ and the right moving one $\phi _{R}$ \ defined in ($%
-L/2$, $\ 0$), ($0$, $L/2$) respectively, in the folded description.
Furthermore to make a connection with the TM we consider more general gluing
conditions:

$\phi _{L}(x=0)=\mp\phi_{R}(x=0)-\varphi _{0}$

the $-$($+$) sign staying for the twisted (untwisted) sector. We are then
allowed to use the boundary states given in \cite{Affleck} for the $c=1$
orbifold at the Ising$^{2}$ radius. The $X$ field, which is even under the
folding procedure, does not suffer any change in boundary conditions and
remains always in the Dirichlet state.

The orbifold nature of this CFT allows us to make contact with the BCFT as
described in \cite{Affleck} for the $m=2$ case. In this case the defect
lines in the Ising model can be seen as BS in the folded model (i.e. the $c=1
$ orbifold or equivalently Ising$^{2}$). They are given in terms of the
Ising ones as:
\begin{eqnarray}
|\uparrow >&=&\frac{1}{\sqrt{2}}\left( ||I>>+||\epsilon >>\right)
+\frac{1}{ 2^{1/4}}||\sigma >> \\ |\downarrow > &=&\frac{1}{\sqrt{2}}\left(
||I>>+||\epsilon >>\right) -\frac{1 }{2^{1/4}}||\sigma >> \\
|f>&=&||I>>-||\epsilon >>
\end{eqnarray}
where $||I>>$, $||\epsilon >>$ and $||\sigma >>$ are the standard Ishibashi
states of the Ising model \cite{cft}. The partition function of an Ising
model with defects has been considered in \cite{Petkova}. There are three
independent possibilities:
\begin{eqnarray}
Z_{0,\frac{1}{2}} &=&(\bar{\chi}_{0}\chi _{\frac{1}{2}}+c.c.)+\bar{\chi}_{%
\frac{1}{16}}\chi _{\frac{1}{16}} \\
Z_{\frac{1}{16},0} &=&\bar{\chi}_{\frac{1}{16}}(\chi _{0}+\chi _{\frac{1}{2}%
})+c.c \\
Z_{\frac{1}{16},\frac{1}{16}} &=&Z_{0,\frac{1}{2}}+Z_{\frac{1}{16},0}
\end{eqnarray}

In the orbifold model there are nine BS given by the product of the two
Ising BS and two continuous ones which cannot be related to the Ising model.
These boundaries fall into $3$ classes with different boundary entropy $%
g_{I}=1$, $1/2$, $1/\sqrt{2}$ respectively and their stability under the
boundary perturbation is obtained according to the ``$g$ theorem'', with $g$
decreasing along a renormalization group trajectory connecting two
conformally invariant boundary conditions. Nevertheless, while in \cite
{Affleck} the two Ising are equivalent in our case the coupling to the
charged sector of the even Ising breaks this $Z_{2}$ symmetry.

The Ishibashi states for the TM are easily obtained as combinations of the
BS of the charged and neutral sector.

The most convenient representation of such BS is the one in which they
appear as a product of Ising and MR BS:
\begin{align}
|\chi _{(0,s)}^{MR}& >=|2s>\otimes |\uparrow >+|2(s+p)+2>\otimes |\downarrow
> \\
|\chi _{(1,s)}^{MR}& >=\frac{1}{2^{1/4}}\left( |2s+1>+|2(s+p)+3>\right)
\otimes |f> \\
|\chi _{(2,s)}^{MR}& >=|2s>\otimes |\downarrow >+|2(s+p)+2>\otimes |\uparrow
>
\end{align}
Such a factorization naturally arises already for the TM characters \cite
{cgm4}. In the above equation the MR BS are given in terms of the BS $%
|\alpha >$ for the charged boson (see ref.\cite{cft} for details) and the
Ising ones.

The vacuum state for the TM model corresponds to the $\tilde{\chi}_{(0,0)}$
character which is the product of the vacuum state for the MR model and that
of the Ising one. As we can see in eqs.(\ref{vacuum1},\ref{vacuum2}) the
lowest energy state appears in two characters. As it has been already shown
that is a characteristics of the orbifold construction and a linear
combination of them must be taken in order to define a unique vacuum state
(see ref.\cite{Cappelli1} for a generalization of the Cardy condition). So
the correct BS in the untwisted sector are:
\begin{align}
|\tilde{\chi}_{((0,s),0)}& >=\frac{1}{\sqrt{2}}\left( |\tilde{\chi}%
_{(0,s)}^{+}>+|\tilde{\chi}_{(0,s)}^{-}>\right) =\sqrt{2}(|2s>\otimes
|\uparrow \bar{\uparrow}>+|2(s+p)+2>\otimes |\downarrow \bar{\uparrow}>)
\label{boud1} \\
|\tilde{\chi}_{((0,s),1)}& >=\frac{1}{\sqrt{2}}\left( |\tilde{\chi}%
_{(0,s)}^{+}>-|\tilde{\chi}_{(0,s)}^{-}>\right) =\sqrt{2}(|2s>\otimes
|\downarrow \bar{\downarrow}>+|2(s+p)+2>\otimes |\uparrow \bar{\downarrow}>)
\\
|\tilde{\chi}_{((1,s),0)}& >=\frac{1}{\sqrt{2}}\left( |\tilde{\chi}%
_{(1,s)}^{+}>+|\tilde{\chi}_{(1,s)}^{-}>\right) =\sqrt{2}(|2s>\otimes
|\downarrow \bar{\uparrow}>+|2(s+p)+2>\otimes |\uparrow \bar{\uparrow}>) \\
|\tilde{\chi}_{((1,s),1)}& >=\frac{1}{\sqrt{2}}\left( |\tilde{\chi}%
_{(1,s)}^{+}>-|\tilde{\chi}_{(1,s)}^{-}>\right) =\sqrt{2}(|2s>\otimes
|\uparrow \bar{\downarrow}>+|2(s+p)+2>\otimes |\downarrow \bar{\downarrow}>)
\\
|\tilde{\chi}_{(s)}(\varphi _{0})& >=\frac{1}{2^{1/4}}\left(
|2s+1>+|2(s+p)+3>\right) \otimes |D_{0}(\varphi _{0})>  \label{continous}
\end{align}
where we also added the states $|\tilde{\chi}_{(s)}(\varphi _{0})>$ in which
$|D_{0}(\varphi _{0})>$ is the continuous orbifold Dirichlet boundary state
defined in ref.\cite{Affleck}. For the special $\varphi _{0}=\pi /2$ value
one obtains:
\begin{equation}
|\tilde{\chi}_{(s)}>=\frac{1}{2^{1/4}}\left( |2s+1>+|2(s+p)+3>\right)
\otimes |ff>
\end{equation}
We briefly discuss the relevance of such a continuous state in sec.(5).

For the twisted sector we have:
\begin{align}
|\chi _{(0,s)}^{+}>& =\left( |2s>+|2(s+p)+2>\right) \otimes (|\uparrow \bar{f%
}>+|\downarrow \bar{f}>) \\
|\chi _{(1,s)}^{+}>& =\frac{1}{2^{1/4}}\left( |2s+1>+|2(s+p)+3>\right)
\otimes (|f\bar{\uparrow}>+|f\bar{\downarrow}>)  \label{boud2}
\end{align}

On the other hand the modular transformations for the characters $\chi
_{(i,s)}^{-}$ depend on the parity of $p$, and as a consequence the BS also
depends on it:

for $p$ even
\begin{eqnarray}
|\chi _{(0,s)}^{-} &>&=\left( |2s>-|2(s+p)+2>\right) \otimes (|\uparrow \bar{%
f}>-|\downarrow \bar{f}>) \\
|\chi _{(1,s)}^{-} &>&=\frac{1}{2^{1/4}}\left( |2s+1>+|2(s+p)+3>\right)
\otimes (|f\bar{\uparrow}>-|f\bar{\downarrow}>)
\end{eqnarray}

for $p$ odd
\begin{eqnarray}
|\chi _{(0,s)}^{-} &>&=\left( |2s>+|2(s+p)+2>\right) \otimes (|\uparrow \bar{%
f}>-|\downarrow \bar{f}>) \\
|\chi _{(1,s)}^{-} &>&=\left( |2s+1>-|2(s+p)+3>\right) \otimes (|f\bar{%
\uparrow}>-|f\bar{\downarrow}>)
\end{eqnarray}

In order to compare the relative stability of the previous boundary states,
let us define the $g$ entropy function for the TM. In general, under
boundary RG flow, the central charge $c$ stays fixed. The flow takes place
in the space of boundary conditions of the bulk theory and a quantity
analogous to $c$ can be defined, that is the boundary entropy $g$ \cite
{Affleck2}. It measures the ``number of boundary degrees of freedom'' and
decreases along the RG flow. Such a quantity can be defined as the term in
the annulus partition function which is independent of the width $L$ of the
strip in the thermodynamic $L$ $\rightarrow \infty $ limit or, equivalently,
as the disk partition function $g_{B}=<0||B>$ of the boundary state $|B>$
associated to the perturbed theory. In our case it can be explicitly
evaluated as $g_{a}=<\tilde{\chi}_{((0,0),0)}||\chi _{a}>$ for any boundary
state $|\chi _{a}>$. Furthermore these BS satisfy a generalized Cardy
condition \cite{Cappelli1} and the $g$ function is expressed as $g=g_{MR}g_{%
\bar{I}}$ where $g_{MR}=\sqrt{2(p+1)}\sin \frac{\pi }{4}(\lambda +1)$.

By using the vacuum state given in eq.(\ref{boud1}) we found the following
values of $g$ for the different classes of boundary conditions:
\begin{tabular}{|l|l|}
\hline
TM & $g$ \\ \hline
$|\tilde{\chi}_{((i,s),f)}>$ & $\frac{\sqrt{p+1}}{2}$ \\ \hline
$|\tilde{\chi}_{(s)}>$ & $\sqrt{2(p+1)}$ \\ \hline
$|\chi _{(i,s)}^{\pm }>$ & $\sqrt{\frac{p+1}{2}}$ \\ \hline
\end{tabular}

\section{Twisted BS and stability}

In order to understand more on the stability of the BS just given before, it
is crucial to view our approach within the framework of the Kondo and
related problems. The low $T$ behavior of a system in the presence of an
impurity is described by an effective CFT in which the impurity disappears
(it is screened) but certain interactions between the quasi-particles are
generated in the screening process. We first discuss the class of impurities
in the even interaction channel which produces a two-channel Kondo like
interaction. In this case only the symmetric boundary terms give
contribution to the dynamics. In general we can introduce two parameters $%
V_{1}$ and $V_{2}$ for the boundary potential for the up and down layer
respectively which can be also written in the even (odd) basis as: $V=\left(
V_{1}+V_{2}\right) /2$ ($\bar{V}=\left( V_{1}-V_{2}\right) /2$). The
relevant quantum number $\lambda _{I}$ of the impurity gets hybridized with
that of the fluid (see the definition of $\lambda $\ in sec.(3)), as it
appears in the boundary state, $\lambda _{I}$ being equivalent to the $su(2)$
spin of the Kondo problem \footnote{%
Notice that the quantum number $s$ defined in sec.(3), describing flux
addition, plays no role in the stability of the ground state and then can be
taken equal to zero. In such a case a diagonal form of the partition
function is recovered.}. The $\lambda _{I}=0$ value corresponds to the free
case (no interactions between the Hall fluid electrons and the impurity),
while $\lambda _{I}=1$ and $\lambda _{I}=2$ correspond to the overscreened
and the exactly-screened Kondo interaction respectively. From the QHE point
of view $\lambda _{I}$ describes a paired ($\lambda _{I}$ even) or unpaired (%
$\lambda _{I}$ odd) impurity. An impurity with $\lambda _{I}=1$ induces a
breaking of the pairing symmetry and a flow which changes the boundary
conditions. Indeed our description adapts very closely to a system of two
interacting Luttinger liquids coupled resonantly through an impurity placed
in between. Such a system is described by the tunneling term:
\begin{equation}
H_{V}=V\left( \cos \alpha X(0)\cos \phi (0)d_{x}^{\lambda _{I}}+\sin \alpha
X(0)\cos \phi (0)d_{y}^{\lambda _{I}}\right) -\tilde{V}d_{z}^{\lambda
_{I}}\partial X(0)  \label{anisKondo}
\end{equation}
where $\alpha =\sqrt{2/(p+1)}$ and $V$, $\tilde{V}$ $=\left( \tilde{V}_{1}+%
\tilde{V}_{2\text{ }}\right) /2$ \ are coupling constants and $%
d_{x}^{\lambda _{I}},d_{y}^{\lambda _{I}},d_{z}^{\lambda _{I}}$ ($\lambda
_{I}=1$ here) describe the impurity ``spin'' placed at the origin. The above
Hamiltonian is similar to that of the closely related problem of a resonant
tunneling junction between quantum wires or hopping of electrons in Quantum
Hall bars coupled to a dot or anti-dot (see ref.\cite{Lin} for details). In
the framework of the Kondo problem such an interaction term is equivalent to
the anisotropic two-channel one impurity Hamiltonian \cite{Lin} and for $V=%
\tilde{V}$ it reduces to the isotropic case. The $V=0,\pm \infty $ conformal
fixed points are unstable and flow to an intermediate stable point $V^{\ast }
$. For the isotropic Kondo problem, i.e. $p=0$, and $m$ channels it is
always possible to ``complete the square'' at the special value $V^{\ast }=%
\frac{2}{2+m}$, where the complete Hamiltonian reduces to its free form
after a shift of the current operators by ${\bold d}^{\lambda_{I}}$ which
preserves the $su(2)_{m}$ Kac-Moody algebra. That corresponds to the
``absorption'' of the impurity and in such a case the system renormalizes
to the intermediate fixed point $V^{\ast }$(``fusion rules
hypothesis''\cite{Affleck2}).

In the more general $V\neq \tilde{V}$ case instead, by performing the
rotation $U=e^{i\tilde{V}d_{z}^{\lambda _{I}}X(0)}\sigma $, ($\sigma $ being
the twist operator of the Ising model) $H_{V}$ reduces to $H_{V}=V_{eff}\cos
\phi (0)d_{x}^{\lambda _{I}}-\tilde{V}d_{z}^{\lambda _{I}}\partial X(0)$.
After such a transformation the boundary conditions for both the charged and
neutral bosons are changed according to the ``fusion principle'' (see ref.
\cite{Affleck}). The $X$ boson acquires a phase shift at the impurity
location while the neutral one gets both twisted and shifted. These new
boundary conditions are consistent with the parity rule for the bulk CFT\
discussed in \cite{cgm4}. The boundary equation of motion can be obtained by
varying the action with respect to all the fields, including $d_{x}^{\lambda
_{I}}$,$d_{y}^{\lambda _{I}}$, which are dynamical variables. The vacuum
states corresponding to these fixed points belong to the twisted sector as
it can be easily understood by noticing that the $U$ operator corresponds to
the $\lambda =1$ representation of the MR model. The total dimension of $U$
is: $h=\frac{1}{8(p+1)}+\frac{1}{16}=\frac{3+p}{16(p+1)}$. At the Toulouse
point the last term cancels out with a term generated by the rotation $U$ in
the free Hamiltonian obtaining $H_{V}=V_{eff}\cos \phi (0)d_{x}^{\lambda
_{I}}$, with the resulting decoupling of the charged field $X$ from the
neutral field $\phi $. Notice that the above transformation changes the
reference state from $<\tilde{\chi}_{((0,0),0)}|$ to $<\chi _{(1,0)}^{+}|$
(see eqs.(\ref{boud1},\ref{boud2})) and at the same time the bare coupling
constant ($V$) appearing in eq.(\ref{anisKondo}) is mapped to the new
effective one ($V_{eff}$). In fact from the RG equation for the boundary
entropy a stable fix point is reached at a finite value of $V$ which is
mapped into an infinite $V_{eff}$ fixed point after the $U$ rotation. As for
the isotropic case, the new interaction becomes that of a $\lambda _{I}=1$
exactly-screened problem.

In particular to the $V_{eff}=0$ value corresponds the unstable boundary
state $|\tilde{\chi}_{(s)}>$ with $g=\sqrt{2(p+1)}$ while to the $%
V_{eff}=\pm \infty $ limit the stable $|\chi _{(1,s)}^{+}>$ ones with $g=
\sqrt{\frac{p+1}{2}}$. In such ground states only one component of the
impurity enters the Hamiltonian while the other one describes non dynamical
fermionic degrees of freedom. That describes a non-Fermi liquid fixed point
for the neutral sector which generalizes the $p=0$ case.

\begin{figure}[p]
\centerline{\epsfxsize=.5cm} \epsffile{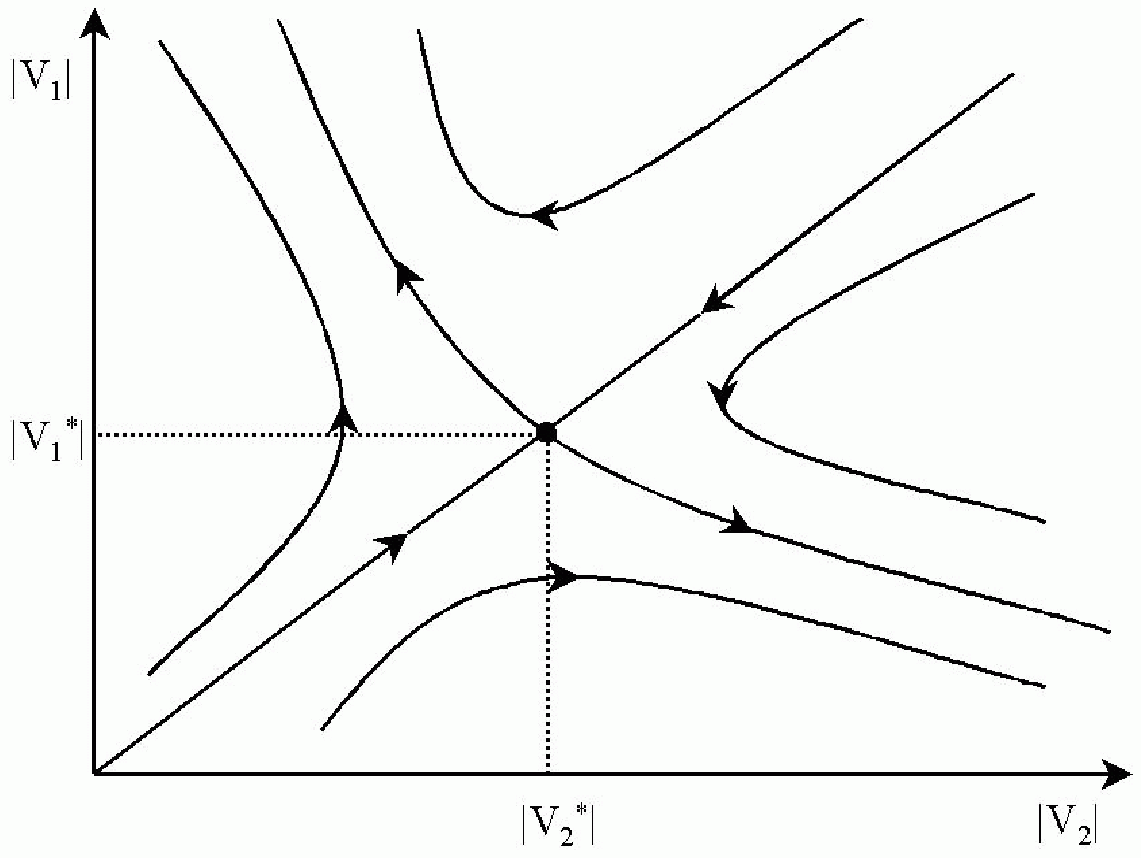}
\caption{A schematic RG flow diagram for the TM in the $V_{1}$, $V_{2}$ parameters space.}
\label{fig:flow}
\end{figure}

As it is well known for the Kondo problem, under a flavour symmetry
breaking perturbation the fixed point is unstable \cite{Kane}. An
asymmetric local tunneling term in the Hamiltonian induces a flow to one of
the two stable points in which the impurity hybridizes with one channel
only. The same structure exists in the present system. A schematic RG flow
in terms of the two coupling constants $V_{1}$, $V_{2}$ of the layers is
given in fig:(\ref {fig:flow}).

The new perturbation gives rise to a flow from the unstable intermediate
fixed point $V^{\ast }$ to the states given by $|\tilde{\chi}_{(0,s)}>$, $|%
\tilde{\chi}_{(1,s)>},$ which are the most stable ones with a value of $g$
given by $g=\frac{\sqrt{p+1}}{2}$. That is close to the case of a
one-channel Kondo problem for the layer interacting with the impurity while
the other one remains completely free. We observe that no quasi-holes are
present in the stable state and by taking a linear combination it is easy to
see that there is a complete factorization of the charged and neutral
degrees of freedom, that is Abelian statistics is recovered!

It is worthwhile to comment briefly on the cocycles here, which have to be
handled with great care in problems with several fermion species. In
addition to the exponential of a free boson, each fermion requires a
cocycle. We also notice that the Kondo analysis should be restricted to the
fermionic case (i.e. for $p$ odd) for which the Pauli principle is at work.
However that does not modify our analysis.

\section{Summary and comments}

As it has been seen in this letter, the possibility of viewing the TM
degrees of freedom in terms of BS has allowed us to establish a close
relation of the twisted theory, previously used for describing a Quantum
Hall system of two interacting layers, with the two-channel one impurity
Kondo problem. In such a context the different possibilities for the bulk
electrons of  exactly-screening or overscreening the impurity spin have been
identified. On the other hand such an identification has allowed us to
understand the degree of stability of the different TM boundary states. In
particular it has been shown that the two layers resonant case corresponds
to the situation in which the impurity spin couples with the same strength
to the electrons of the two different channels, i.e. the layers in the
Quantum Hall bilayer ( $V_{1}=V_{2}$). Then the two layers are completely
balanced and the ground state shows non-Abelian statistics due to the
existence of quasi-holes states. From the analysis of the flow of the
boundary entropy $g$ such a vacuum is not stable under antisymmetric
perturbations (it is a saddle point). For strong coupling of the impurity
spin with the bulk electrons belonging to only one of the two layers ($%
|V_{1}|$ or $|V_{2}|\rightarrow \infty $) the system degenerates towards the
most stable ground state with no quasi-holes to condense, the degrees of
freedom left being only electrons and anyons states with Abelian statistics.
Therefore we conjecture that non-Abelian statistics can be realized only in
an extremely clean sample while in the presence of impurities the statistics
reduces to the Abelian one. Furthermore the twist at the origin couples the
two layers in a topological way, with a consequent phase shift between
electrons in the up and down layer which is constant. Such a phase
difference fixed to $\pi $ for the twisted sector ground state, can be
deformed to a continuous value $\varphi _{0}$ in the boundary state $|\tilde{%
\chi}_{(s)}(\varphi _{0})>$ allowing for a Josephson like effect, which is
at the moment under study. Another interesting problem to study is the case
in which there are both boundary and bulk perturbations. In this case the
flow can act also on the central charge of the CFT and the reduction to a
pure MR model could be realized. Due to the analogy with the string and
D-brane dynamics, the present analysis also applies to D-branes systems as
it was done in \cite{cgm3}. Finally we notice that for an impurity with
quantum number $s_{I}\neq 0$ the scattered particle might change statistics
and at the same time the partition function would be in a non-diagonal
form.

\bigskip

{\bf Acknowledgments} - We thank M. Huerta, A. Cappelli and G. Mussardo for
useful comments and for reading the manuscript.

\pagebreak

\end{document}